\documentclass{article}

\usepackage{setspace}
\usepackage[english]{babel}
\usepackage[utf8]{inputenc}
\usepackage[T1]{fontenc}
\usepackage{multimedia}
\usepackage{amsfonts,multirow}
\usepackage{amsmath,bm}
\usepackage{amsthm}
\usepackage{amssymb}
\usepackage{epsfig}
\usepackage{aeguill}
\usepackage{graphicx,psfrag,color}
\usepackage{subfigure}
\usepackage{times}

\usepackage[superscript,biblabel]{cite}
\usepackage[super,comma]{natbib}

\renewcommand{\P}{{\rm I}\kern-0.14em{\rm P}}
\newcommand{\1}{{\rm 1}\kern-0.28em{\rm I}}
\newcommand{\E}{{\rm I}\kern-0.14em{\rm E}}
\newcommand{\R}{{\rm I}\kern-0.14em{\rm R}}
\newcommand{\reals}{{\rm I}\kern-0.16em{\rm R}}

\newcommand{\p}{{\rm I}\kern-0.18em{\rm P}}

\newcommand{\eps}{{\varepsilon}}

\newcommand{\beq}{\begin{equation}}
\newcommand{\eeq}{\end{equation}}

\makeatletter
\newcommand*{\indep}{%
  \mathbin{%
    \mathpalette{\@indep}{}%
  }%
}
\newcommand*{\nindep}{%
  \mathbin{
    \mathpalette{\@indep}{\not}
  }%
}
\newcommand*{\@indep}[2]{%
  \sbox0{$#1\perp\m@th$}
  \sbox2{$#1=$}
  \sbox4{$#1\vcenter{}$}
  \rlap{\copy0}
  \dimen@=\dimexpr\ht2-\ht4-.2pt\relax
  \kern\dimen@
  {#2}%
  \kern\dimen@
  \copy0 
} 
\makeatother

\usepackage{tikz}
\usetikzlibrary{arrows,shapes}

\tikzset{
semi/.style={
  semicircle,
  draw,
  minimum size=2em
  }
}

\tikzstyle{var}=[draw,circle,thick, text width=4mm, inner sep = 3pt]
\tikzstyle{varr}=[draw,circle,thick]
\tikzstyle{varf}=[draw=none,fill=none]
\tikzstyle{varCond}=[draw,rectangle,rounded corners=3pt, fill=gray, minimum width=2em,minimum height=2em]

\tikzstyle{varcounterfactual}=[draw, ellipse] 
\tikzstyle{varright} = [draw, semi, shape border rotate=270] 
\tikzstyle{varleft} = [draw,  semi,shape border rotate=90]

\tikzstyle{edge} = [draw,thick,->,>=latex]
\tikzstyle{edge2} = [draw,thick,-]
\tikzstyle{edge3} = [draw,dashed,->,>=latex]

\title{{\bf Can collider bias fully explain the obesity paradox?}}

\author{Vivian Viallon$^{*(1)}$ and Marine Dufournet$^{(2)}$\\ $^{(1)}$Univ Lyon, Universit\'e Lyon 1, IFSTTAR, UMRESTTE, F-69373 Lyon\\
$^{(2)}$IFSTTAR, Univ Lyon, Universit\'e Lyon 1, UMRESTTE, F-69373 Lyon\\
$^*$ vivian.viallon@univ-lyon1.fr}
\date{}

\begin{document}
\maketitle

\section*{Abstract }

\paragraph{Background: }
The ``obesity paradox'' has been reported in several observational studies, where obesity was shown to be associated to a decreased mortality in individuals suffering from a chronic disease, such as diabetes or heart failure. Causal arguments have recently been given to explain this apparently paradoxical fact: because the chronic disease is caused by obesity, the observed ``protective effect'' of obesity among patients with, say, diabetes, actually has no causal value. Recently, Sperrin et al.  \cite{Sperrin} relaunched the debate and claimed that the resulting bias, the so-called collider bias, was unlikely  to  be  the  main  explanation for the obesity paradox. However, a number of issues in their work make their conclusions questionable.

\paragraph{Methods :} We first study the bias between $(i)$ the  association between obesity and early death among patients suffering from the chronic disease $\Delta_{AS}$ and $(ii)$ the causal effect considered by Sperrin et al.. Under the usual framework of structural causal models, we explain why this bias  can be much higher than what these authors reported. We further consider alternative causal effects of potential interest and study their difference with $\Delta_{AS}$. Numerical examples are presented to illustrate the magnitude of these differences under realistic scenarios. 

\paragraph{Results : } We show that it is possible to have a negative  $\Delta_{AS}$, while the causal effects we considered are all positive. 

\paragraph{Conclusion: } Even under the very simple generative model we considered, collider bias can be the sole cause of the ``obesity paradox''. 
\newpage

\section{Introduction}

The ``obesity paradox'' has been reported in several observational studies, where obesity was shown to be associated to a decreased mortality in individuals suffering from a chronic disease, such as diabetes \cite{Carnethon} or heart failure \cite{Oreopoulos}. Some biological explanations have been put forward, but causal arguments have recently been given to explain this apparently paradoxical fact \cite{Lajous_2014}: because the chronic disease is on a causal path between  obesity and mortality, conditioning on the value of this mediator can create spurious association, or bias. And the observed ``protective effect'' of obesity among patients with, say, diabetes, actually has no causal value.

Recently, Sperrin et al.  \cite{Sperrin} relaunch the debate and claim that the resulting bias, the so-called collider bias, is unlikely  to  be  the  main  explanation for the obesity paradox. However, we believe there are a number of issues in Sperrin et al.'s article which make the authors' conclusions questionable. We further believe that a thorough description of these issues can be instructive for epidemiologists interested in causal inference. 

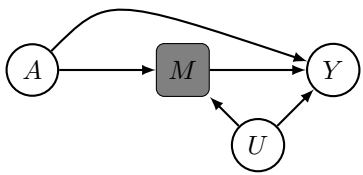
\begin{figure}[t]
\caption{The DAG considered in \cite{Sperrin}} 
\label{fig:DAG} 
\vspace{0.3cm}
\begin{center}
\begin{tikzpicture}[scale=1, auto,swap]
\node[varr] (A)at(0,0){$A$};
\node[varr] (Y)at(4,0){$Y$};
\node[varCond] (M)at(2,0){$M$};
\node[varr] (U) at (3,-1) {$U$};
\draw[edge] (U)--(M);
\draw[edge] (U)--(Y);
\draw[edge] (A)--(M);
\draw[edge] (M)--(Y);
\draw[edge] (A).. controls (1,1) ..(Y);
\end{tikzpicture}
\end{center}
\end{figure}

Sperrin et al. based their arguments on a simple form of the causal model underlying the obesity paradox; see Figure \ref{fig:DAG}. The four random variables $A$, $M$, $Y$ and $U$ are binary variables, {\em e.g.} corresponding to obesity, diabetes, early death and some (potentially unobserved) binary confounder, respectively. 
For simplicity, the authors first consider effects and associations defined on the additive scale. In particular, the association between $A$ and $Y$ given $M=1$ is  
$$\Delta_{AS} = E[Y|M=1, A=1] - E[Y| M=1, A=0]. $$
This quantity can be estimated from data collected from patients with diabetes ($M=1$), even if $U$ is unobserved. As mentioned above, several observational studies reported negative estimates for $\Delta_{AS}$, leading to the obesity paradox. However, $\Delta_{AS}$ has to be related to a meaningful causal effect for the paradox to be real. Sperrin et al. then study the bias between $\Delta_{AS}$ and the causal effect of $A$ on $Y$ conditioned on $M=1$, which they define as 
$$ \Delta_{CE} = E[Y^{A=1}| M=1] - E[Y^{A=0}| M=1].$$
Here, $Y^{A=1}$ and $Y^{A=0}$ are two counterfactual variables or potential outcomes, that one would have been able to observe in the counterfactual worlds $\Omega^{A=1}$ and $\Omega^{A=0}$ that would have followed the intervention $do(A=1)$ and $do(A=0)$, respectively \cite{Pearl_2000}.  See Section \ref{sec:Details} for the precise definitions of $Y^{A=a}$, $a\in\{0, 1\}$, under the model of Figure \ref{fig:DAG}.  
A first remark is that there is no unique and well-defined intervention $do(A=1)$ or $do(A=0)$ when $A$ represents obesity. As nicely put forwarded in \cite{Hernan_2008}, this makes causal inference about obesity a particularly difficult task, because assumptions such as consistency, positivity and exchangeability are unlikely to hold \cite{Hernan_2008}.  Here we ignore this problem just as Sperrin et al. did: we assume the existence of such an intervention and proceed as usual under structural causal models \cite{Pearl_2009, Hernan_Robins_Book}.


Along their derivation, it seems that Sperrin et al. assume that $Y^{A=a}$ is independent of $A$ given $M=1$ whereas  this is generally not the case under the model of Figure \ref{fig:DAG}, because $M$ is a descendant of $A$. Then, they obtain an incorrect expression for $\Delta_{CE}$ and the true bias $\Delta_{CE} - \Delta_{AS}$ is generally much larger than what they report; see Section \ref{sec:Error} below. As will be shown in Section \ref{sec:Simul}, many configurations of the simple generative model considered by Sperrin et al. lead to an "observed protective effect" of obesity among diabetic patients $(\Delta_{AS} < 0)$ while the causal effect $\Delta_{CE}>0$ is positive. Stated another way, $(\Delta_{AS} < 0)$ does not imply $(\Delta_{CE} < 0)$ under the model considered by Sperrin et al.: if the expression ``obesity paradox'' is intended to mean that the causal effect of obesity on early death, among the patients with chronic disease {\em in the actual world}, is negative, $\Delta_{CE} <0$, then there is no way to conclude that this obesity paradox is real from studies reporting a negative value for (an estimate of) $\Delta_{AS}$, because of collider bias.

In other respect, the premise that we are interested in $\Delta_{CE}$ is questionable. Indeed, $\Delta_{CE}$ relies on the ``cross-world'' quantities $E[Y^{A=a}| M =1]$ and its interpretation is not straightforward (see Section \ref{sec:Error_remarks}). Because $M$ is a mediator here, alternative causal effects of interest can be found in the literature dealing with mediation analysis. Denoting by $Y^{A=a,  M=m}$  the outcome variable we would have been able to observe in the counterfactual world $\Omega^{a,m}$ that would have followed the double intervention $do(A=a, M=m)$, the controlled direct effect at $M=1$ is defined as  $\Delta_{CDE}=E[Y^{A=1,  M=1}] - E[Y^{A=0, M=1}]$. Conditional versions of $\Delta_{CDE}$ can be considered as well (see Section \ref{sec:Misconception}). We show that the bias  $(\Delta_{CDE} - \Delta_{AS})$ is typically lower than $(\Delta_{CE} - \Delta_{AS})$, and is close to what Sperrin et al. incorrectly described as the bias $(\Delta_{CE} - \Delta_{AS})$. This suggests that if the obesity paradox means that the controlled direct effet $\Delta_{CDE}$ is negative, then Sperrin et al.'s discussion, when applied to the bias between the estimable quantity $\Delta_{AS}$ and the target quantity $\Delta_{CDE}$, is mostly valid. In Section \ref{sec:Simul}, we present numerical results obtained under the same data generation mechanism as Sperrin et al., but we eventually consider additional interaction terms. For some particular configurations, we obtain $\Delta_{AS}<0$ and $\Delta_{CDE}>0$. Therefore,  a negative $\Delta_{AS}$ implies neither a negative $\Delta_{CE}$ nor a negative $\Delta_{CDE}$ and collider bias can fully explain the ``apparent paradox''.

\section{Identifiability of $\Delta_{CE}$ and study of $\Delta_{CE} - \Delta_{AS}$}\label{sec:Error}

\subsection{General presentation}
Recall the definition $\Delta_{CE} = E[Y^{A=1}| M=1] - E[Y^{A=0}| M=1]$. Because it relies on counterfactual variables, Sperrin et al.  tried to express it in terms of the observed variables $A, M, U$ and $Y$. But, they incorrectly establish that   $\Delta_{CE}=\Delta_{Sp}$, with $\Delta_{Sp}$ defined as: 
$$ \sum_u  \left\{E[Y | M=m, A=1, U=u] - E[Y | M=m, A=0, U=u] \right\}P[U=u|M=m].$$
On the left column of Page 526, the authors especially write 
\begin{align}
E[Y^{A=a}| M=m] &= \sum_u  E[Y^{A=a} | M=m, U=u]P[U=u|M=m]\nonumber \\
&=\sum_u  E[Y | M=m, A=a, U=u]P[U=u|M=m]. \label{quant_err}
\end{align}
But Equality (\ref{quant_err}) is only guaranteed if $E[Y^{A=a} | M=m, U=u]=E[Y | M=m, A=a, U=u]$ which, on turn, is only guaranteed if $Y^{A=a}\indep A \mid (M, U)$. However, this conditional independence does generally not hold under the model depicted in Figure \ref{fig:DAG} because 
$M$ is a descendant of $A$ and, then, the set $\{M,U\}$ does not satisfy the back-door criterion; see \cite{Pearl_2000, Pearl_2010}. See also Fine point 7.2 in \cite{Hernan_Robins_Book} where the authors use the SWIG approach \cite{SWIGs} in a related causal model. As a result, $\Delta_{Sp}$ is generally different from $\Delta_{CE}$ under the causal model of Figure \ref{fig:DAG}. 

As a matter of fact, $\Delta_{CE}$  can not be expressed in terms of the distribution of the variables $(A, M, U, Y)$ without further assumptions on the causal model. In Section \ref{sec:Simul} below, we consider a generative model that is consistent with that considered by Sperrin et al. By specifying  the structural functions $f_Y, f_M, f_A$ and $f_U$ and the distributions of the disturbances $\eps_Y$, $\eps_A$, $\eps_M$ and $\eps_U$ in this model, an analytic formula for $\Delta_{CE}$ can be derived (see Section \ref{sec:Details_CE}). Then, we show that the difference between the true value of $\Delta_{CE}$ and $\Delta_{Sp}$, as well as the bias $\Delta_{CE}-\Delta_{AS}$, can be sensible, which invalidates Sperrin et al.'s conclusions.

\subsection{Additional remarks}\label{sec:Error_remarks}

\begin{figure}[t]
\caption{A simplified version of the DAG considered in Sperrin et al., corresponding to the special case of no confounder and no direct effect of $A$ on $Y$.} 
\label{fig:DAG_simple} 
\begin{center}
\begin{tikzpicture}[scale=1, auto,swap]
\node[varr] (A)at(0,0){$A$};
\node[varr] (Y)at(4,0){$Y$};
\node[varCond] (M)at(2,0){$M$};
\phantom{\node[var] (U) at (3,-1) {$U$};}
\draw[edge] (A)--(M);
\draw[edge] (M)--(Y);
\phantom{\draw[edge] (A).. controls (1,-1) ..(Y);}
\end{tikzpicture}
\end{center}
\end{figure}
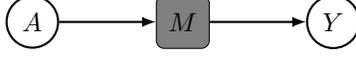

The quantity $\Delta_{CE} = P[Y^{A=1}=1|M=1] - P[Y^{A=0}=1|M=1]$ has to be interpreted with caution. Denote by $M^{A=a}$ the counterfactual variable pertaining to the chronic disease that we would be have been able to observe in the counterfactual world $\Omega^{A=a}$. Further introduce $Y^{a,m} = Y^{A=a, M=m}$ the counterfactual outcome variable that we would have observed in the counterfactual world $\Omega^{a,m}$ that would have followed the double intervention $do(A=a, M=m)$. Observe that $Y^{A=a} = Y^{A=a, M^{A=a}}$ (see Section \ref{sec:Details} for more details), and then that $P[Y^{A=a} = 1 | M=1]  = P[Y^{A=a, M^{A=a}} = 1| M=1]$. In particular, $P[Y^{A=0} = 1 | M=1] = P[Y^{A=0, M^{A=0}} = 1| M=1]$ is the risk of early death in $\Omega^{A=0}$, for the individuals with diabetes in the actual world.  But $M=1 \nRightarrow M^{A=0} = 1$: some diabetic patients in the actual world would not have been diabetic in $\Omega^{A=0}$. As a result, a positive $\Delta_{CE}$, for instance, can be solely due to the fact that a portion of the individuals with diabetes in the actual world would not have suffered from diabetes in $\Omega^{A=0}$ and would have lived longer in $\Omega^{A=0}$ than they would have in $\Omega^{A=1}$. In other words, even if it is conditioned on $M=1$,  $\Delta_{CE}$ is not related to the direct effect of $A$ on $Y$. It is the total effect of $A$ among a subgroup of the population defined according to a variable observed in the actual world, just as the average causal effect on the treated $E[Y^{A=1}-Y^{A=0}| A=1]$ is a measure of the total effect of $A$ on another subgroup of the population. In particular, even if $A$ has no direct effect on $Y$, that is if its effect is entirely mediated through $M$, $\Delta_{CE}$ is generally non-null, contrary to what Sperrin et al. wrote on the right column of Page 526. More precisely, they consider the case where $\beta_A = \beta_{AM} = 0$ under their generative model, which corresponds to a situation where $A$ has no direct effect on $Y$. Then, they claim that $P[Y^{A=1} =1 |M=1] = P[Y^{A=0} =1 |M=1]$, so that $\Delta_{CE}=0$. As explained above, this is generally false (see Figure \ref{fig:Figure2Sp} in  Section \ref{sec:Simul} for an illustration).

Considering in more details the simple model where $U$ is absent from the DAG and $A$ has no direct effect is instructive ; see Figure \ref{fig:DAG_simple}. In this DAG, the empty set satisfies the back-door criterion and, then, $Y^{A=a}\indep A$. As a result, $P[Y^{A=a}=1] = P[Y=1 | A=a]$, and the average total effect is generally non-null, as expected: 
\begin{equation} 
P[Y^{A=1}=1] - P[Y^{A=0}=1] = P[Y=1 | A=1] - P[Y=1 | A=0] \neq 0. \label{eq:Absurd1}
\end{equation}
More precisely, this total effect is always non-null except in the absence of either the arrow pointing from $A$ to $M$ or the arrow pointing from $M$ to $Y$.   Next, we have $Y\indep A|M$ so that $P[Y =1 |A=1, M=1] = P[Y=1 |A=0, M=1]$ and $\Delta_{AS}=0$. As will be made clearer in Section \ref{sec:Misconception}, this implies that the controlled direct effect $\Delta_{CDE}$ is null, as expected too. However, because $\{M\}$ does not satisfy the back-door criterion, $Y^{A=a} \indep A| M$ generally does not hold and $P[Y^{A=a} =1 |M=1]$ generally differs from $ P[Y=1 |A=a, M=1]$ (and $\Delta_{CE}$ is generally non-null). Under this simple model, the fact that  $Y^{A=a} \nindep A| M$ can also be shown by a simple reduction to absurdity argument. First recall that the consistency assumption ($A=a \Rightarrow Y  =Y^{A=a}$) holds under  the structural causal models considered throughout this article (see Section \ref{sec:Details}). Then, if  $Y^{A=a}$ was independent of  $A$ given  $M$, we would get the following chain of equalities:
\begin{eqnarray*}
P[Y^{A=a} = 1] &=&\sum_m P[Y^{A=a} = 1| M=m]P[M=m]  \quad{\rm by\ the\ tower\ rule}\\ 
&=& \sum_m P[Y^{A=a}| M=m, A=a]P[M=m] \quad{\rm if}\ Y^{A=a} \indep A| M  \\ 
&=& \sum_m P[Y=1 | M=m, A=a]P[M=m] \quad{\rm by\ consistency} \\ 
&=& \sum_m P[Y=1 | M=m]P[M=m], \quad{\rm because}\ Y\indep A|M\\
&=& P[Y=1]  \quad{\rm by\ the\ tower\ rule}.
\end{eqnarray*}
Therefore, the assumption $Y^{A=a} \indep A| M=1$ yields $P[Y^{A=1} = 1] - P[Y^{A=0} = 1] = P[Y=1] - P[Y=1] = 0$, which contradicts Equation (\ref{eq:Absurd1}) .  This completes our reduction to absurdity argument and establishes that $Y^{A=a} \indep A| M=1$ is generally false under the model of Figure \ref{fig:DAG_simple}.

In our chain of equalities above, the fact that $P[Y^{A=a} = 1] = \sum_m P[Y^{A=a}=1| M=m]P[M=m]$ follows from the tower rule, that is $E(Z) = E[E(Z|X)]$ for "any" couple of random variables $X$ and $Z$. Even if $Y^{A=a}$ is not observed, we may recall that it is still a random variable in the usual sense. Indeed, the counterfactual variables and observed variables are random variables and are all defined on a common probability space as deterministic functions of the exogenous variables $\eps_A$, $\eps_M$ and $\eps_Y$; see Section \ref{sec:Details} below as well as Sections 4 and 5 in \cite{Pearl_2009} and Section 3.4 in \cite{Pearl_2010} for instance.  As a result, standard probability calculus, among which the tower rule, apply on counterfactual variables too. 

Lastly, using the SWIG approach \cite{SWIGs} for instance, it is easy to show  that $A \indep (M^{A=a}, Y^{A=a})$ and, then, $Y^{A=a}\indep A | M^{A=a}$ under the model of Figure \ref{fig:DAG_simple}; see Section \ref{sec:Details}. Then, successively applying the facts that  $A=a \Rightarrow (Y=Y^{A=a}$ and $M =M^{A=a})$ and $Y^{A=a}\indep A | M^{A=a}$, it follows that
\begin{align*}
P(Y=1 | A=a, M=1)  &= P[Y^{A=a} = 1| A=a, M^{A=a}=1]\\
&= P[Y^{A=a} = 1| M^{A=a}=1]. 
\end{align*}
This is another way to establish that, generally, $P(Y=1 | A=a, M=1) \neq P[Y^{A=a} = 1| M =1]$ in the model of Figure \ref{fig:DAG_simple}. 
Indeed, the random sets $\{M^{A=a}=1\}$ and $\{M=1\}$ are generally different and so are quantities $P[Y^{A=a} = 1| M^{A=a}=1]$ and $P[Y^{A=a} = 1| M =1]$. For instance, $\{M^{A=0}=1\}$ consists of the individuals who would have suffered from diabetes in the counterfactual world $\Omega^{A=0}$, while $\{M=1\}$ consists of the individuals with diabetes in the actual world, among whom some are obese and others are not. If obesity causes diabetes it is clear that $\{M^{A=1}=1\} \neq \{M^{A=0}=1\}$ and then that $\{M=1\}$ differs from $\{M^{A=a}=1\}$, for $a\in\{0,1\}$.  See Section \ref{sec:Details} for more details on the difference between the random sets $\{M^{A=a}=1\}$ and $\{M=1\}$.

\section{Controlled direct causal effects}\label{sec:Misconception}

As mentioned above, $\Delta_{CE}$ has to be interpreted with caution and is not related to the direct effect of $A$. In the presence of a mediator like $M$ here, alternative causal effects have been advocated \cite{imai_general_2010, vanderweele_odds_2010}. In particular, the control direct effect captures the effect of $A$, while fixing the value of the mediator ({\em e.g.} to 1), and is therefore appealing in the context of the obesity paradox. The three following quantities can be considered:
\begin{align*}
\Delta_{CDE} &= P[Y^{1, 1}=1] - P[Y^{0, 1} = 1]\\
\Delta_{CDE|M=1} &= P[Y^{1, 1}=1| M=1] - P[Y^{0, 1}=1| M=1]\\
\Delta_{CDE|A=1, M=1} &= P[Y^{1, 1}=1| A = 1, M=1] - P[Y^{0, 1}=1| A=1, M=1].
\end{align*}
The first one, $\Delta_{CDE}$, is the controlled direct effect of $A$ on $Y$, at $M=1$ \cite{imai_general_2010, vanderweele_odds_2010}. It compares the risk of death in counterfactual worlds $\Omega^{1,1}$ and $\Omega^{0,1}$, that would have followed the double interventions $do(A=1, M=1)$ and $do(A=0, M=1)$, respectively. All individuals suffer from diabetes in these two counterfactual worlds, but they are all obese in $\Omega^{1,1}$, while none of them is obese in $\Omega^{0,1}$. Therefore, $\Delta_{CDE}$ captures the direct causal effect of obesity, while controlling for the diabetic status ($=1$ here). The other two quantities are conditional versions of $\Delta_{CDE}$. More precisely, $\Delta_{CDE|M=1}$ corresponds to $\Delta_{CDE}$ when focusing on individuals who suffer from diabetes in the actual world, while $\Delta_{CDE|M=1,A=1}$ focuses on individuals who are obese and suffer from diabetes in the actual world.

Under the structural causal model corresponding to Figure \ref{fig:DAG}, both $\Delta_{CDE}$ and $\Delta_{CDE|A=1, M=1}$ can be identified from the distribution of $(A, M, U,Y)$. Moreover, the corresponding formulas exhibit similarities with both $\Delta_{AS}$ and $\Delta_{Sp}$. 
Because there is no unobserved confounder between $A$ and $Y$ (except $U$, potentially),  we have $Y^{a,m}\indep A| U$. \cite{vanderweele_odds_2010} Similarly, because there is no unobserved confounder (except, potientially, $U$) between $M$ and $Y$, we have  and $Y^{a,m}\indep M| (A,U)$. \cite{vanderweele_odds_2010}  Therefore, 
\begin{align*}
P[Y^{a,m} = 1] &= \sum_{u} P[Y^{a,m}=1 | U =u] P[U=u]  \quad{\rm by\ the\ tower\ rule} \\
&= \sum_u  P[Y^{a,m} =1 | A=a, U =u] P[U=u]  \quad{\rm because }\ Y^{a,m}\indep A| U\\
&= \sum_u  P[Y^{a,m}=1 |A=a, M=m, U=u]  P[U=u] \quad{\rm because }\ Y^{a,m}\indep M| (A,U)\\
&= \sum_u  P[Y =1 | A=a, M=m, U=u] P[U=u]  \quad{\rm by\ consistency}.
\end{align*}
Then, $\Delta_{CDE}$ can be written
$$\Delta_{CDE} = \sum_{u} \{P[Y=1 | A=1, M=1, U=u] - P[Y=1 | A=0, M=1, U=u] P(U=u)\}.$$
This expression ensures that $\Delta_{CDE}=0$ in the model depicted in Figure \ref{fig:DAG_simple}.

Now, turning our attention to $\Delta_{CDE|A=1,M=1}$, we have 
\begin{align*}
& P[Y^{a,m} = 1 | A=a^*, M=m^*]  \\
&=  \sum_{u}  P[Y^{a,m} = 1 | A=a^*, M=m^*, U=u] P(U=u| A=a^*, M=m^*)\quad{\rm by\ the\ tower\ rule}\\
&=   \sum_{u} P[Y^{a,m} = 1 | A=a^*,  U=u] P(U=u| A=a^*, M=m^*)\quad{\rm because }\ Y^{a,m}\indep M| (A,U)\\
&=   \sum_{u}  P[Y^{a,m} = 1| A=a,  U=u]P(U=u| A=a^*, M=m^*) \quad{\rm because }\ Y^{a,m}\indep A| U\\
&=   \sum_{u}   P[Y^{a,m} = 1 | A=a, M=m, U=u]P(U=u| A=a^*, M=m^*) \quad{\rm because }\ Y^{a,m}\indep M| (A,U)\\
&=   \sum_{u}  P[Y = 1 | A=a, M=m, U=u]P(U=u| A=a^*, M=m^*) \quad{\rm by\ consistency}.
\end{align*}
Therefore,
\begin{align*}
\Delta_{CDE|A=1,M=1} &= \sum_{u} \{P[Y=1 | A=1, M=1, U=u] \\ & - P[Y=1 | A=0, M=1, U=u]\} P(U=u| A=1, M=1).
\end{align*}
Turning our attention back on $\Delta_{AS}$, observe that it writes
\begin{align*}
\sum_{u} \{P[Y=1 | A=1, M=1, U=u]P(U=u| A=1, M=1) \\
- P[Y=1 | A=0, M=1, U=u] P(U=u| A=0, M=1)\}.
\end{align*}

Sperrin et al. claim that the difference between  $\Delta_{AS}$ and $\Delta_{CE}$ is possible because $P(U=u| A=a, M=m)$ generally differs from $P(U=u|M=m)$. As we showed in Section \ref{sec:Error} above, their claim is only valid for the difference between  $\Delta_{AS}$ and $\Delta_{Sp}$, and the difference between $\Delta_{AS}$ and $\Delta_{CE}$ is also due to the discrepancy between $P(Y=1|A=a, M=1)$ and $P[Y^{A=a}|M=1]$. From the formula above, Sperrin et al.'s discussion is actually valid for the difference between  $\Delta_{AS}$, $\Delta_{Sp}$, $\Delta_{CDE}$ and $\Delta_{CDE|A=1,M=1}$. Indeed, the only differences between these four quantities lie in the version of the $U$-distribution used to marginalize the quantities $P[Y=1| A=a, M=1, U=u]$ over $u$, for $a\in\{0,1\}$. It is $P(U=u|A=a, M=1)$ for $\Delta_{AS}$, $P(U=uM=1)$ for $\Delta_{Sp}$,  $P(U=u)$ for $\Delta_{CDE}$ and $P(U=u| M=1, A=1)$ for $\Delta_{CDE|A=1, M=1}$. Despite these similarities, these four quantities are generally different since $M$ typically depends on both $A$ and $U$ under the model of Figure \ref{fig:DAG}. Moreover, our numerical results show that it is possible to have $\Delta_{AS}<0$ while $\Delta_{CDE}>0$ and $\Delta_{CDE|A=1, M=1}>0$ (see Section \ref{sec:Simul}), contradicting Sperrin et al.'s conclusion (even when applied to the controlled direct effets instead of the $\Delta_{CE}$).

Way may further mention that quantities $\Delta_{CDE|M=1}$ and $\Delta_{Sp}$ are generally different. These two quantities would be equal if $Y^{a,m} \indep A | (M, U)$, but this conditional independence does usually not hold under the model of Figure \ref{fig:DAG}. More generally, we were not able to relate $\Delta_{Sp}$ to any meaningful causal effect under this model. 

A final remark is that  $\Delta_{CDE|M=1}$ can not be identified from the distribution of $(A, M, U,Y)$ under the model of Figure \ref{fig:DAG} without further assumptions. It may be identifiable after specifying the generating functions and the disturbances distributions following the same arguments as those used in the case of $\Delta_{CE}$ (see Section \ref{sec:Details_CE}).



\section{Numerical illustration}\label{sec:Simul}

We now provide a few numerical examples illustrating the differences between the various quantities introduced above. Our objective is to show that a negative $\Delta_{AS}$ is not necessarily ``paradoxical'': more precisely, it does not imply either $\Delta_{CE}<0$ or $\Delta_{CDE}<0$ or $\Delta_{CDE|A=1, M=1}<0$.

\subsection{Data generation mechanism}\label{sec:DataGen}

We consider a generative model that can be seen as a special case of, and is then consistent with, the one described by \cite{Sperrin}. More precisely, our data generation mechanism is obtained by specifying the structural functions $f_Y$, $f_M$, $f_A$ and $f_U$ as well as the distributions of the disturbances $\eps_Y$, $\eps_M$, $\eps_A$ and $\eps_U$, which together lead to the same relationships between $Y$, $A$, $M$ and $U$ as those considered in Sperrin et al.  Keep in mind that we had to specify the causal model in order to derive an analytic formula for $\Delta_{CE}$ (see Section \ref{sec:Details_CE}). 

Denote by $\1[\cdot]$ the indicator function. Define four independent random variables $\eps_A$,  $\eps_U$, $\eps_M $ and $\eps_Y$ distributed according to a uniform distribution over the interval $[0,1]$. For any given  $(p_A, p_U)\in(0,1)^2$, define $A=\1[\eps_A \leq p_A]$ and $U=\1[\eps_U\leq p_U]$ so that $A \sim B(p_A)$ and $U\sim B(p_U)$ are two independent Bernoulli variables ; as in Sperrin et al.'s work, we consider the special case where $p_A=p_U=0.5$. Now, introduce the sigmoid function ${\rm expit}(x) = (1+ \exp(-x))^{-1}$, and set, for any $(a, u, m)\in\{0,1\}^3$ and for some real parameters $\alpha_0, \alpha_A, \alpha_U, \alpha_{AU}, \beta_0, \beta_A, \beta_U, \beta_M, \beta_{AM}, \beta_{AU}, \beta_{UM}$ and $\beta_{AUM}$,
\begin{align*}
p_{M}(a,u) &= {\rm expit}(\alpha_0 + \alpha_A a + \alpha_U u + \alpha_{AU}au) \\ 
p_{Y}(a,m,u) &= {\rm expit}(\beta_0 + \beta_A a + \beta_U u + \beta_M m + \beta_{AU} au +\\
&\quad\quad\quad\quad \beta_{AM} am + \beta_{UM} um + \beta_{AUM} aum).
\end{align*}
Finally, variables $M$ and $Y$ are  defined as
\begin{align*}
M &=f_M(A, U, \eps_M)=\1[\eps_M \leq p_{M}(A,U)], \\
Y &=  f_Y(A, M, U, \eps_Y)= \1[\eps_Y \leq p_{Y}(A,M,U)].
\end{align*}

Sperrin et al. only considered situations where interaction terms in the $Y$-model were all null: $\beta_{AM}=\beta_{AU}=\beta_{UM}=\beta_{AUM}=0$. We will show below that conclusions can be quite different when considering non-zero values for these parameters, especially when comparing $\Delta_{AS}$ and $\Delta_{CDE}$ or $\Delta_{CDE|A=1, M=1}$.  Moreover, following Sperrin et al., we set
\begin{align*} 
\alpha_0 &=  -\frac{1}{2}\left(\alpha_A+ \alpha_U+\frac{1}{2}\alpha_{AU}\right)\\
\beta_0 &=  -\frac{1}{2}\left(\beta_A+\beta_M+\beta_U + \frac{1}{2} (\beta_{AM} + \beta_{AU} + \beta_{UM}) + \frac{1}{4}\beta_{AUM} -\nu \right)
\end{align*}
for some $\nu\geq 0$. Sperrin et al. choose $\nu=0$, which ensures that the overall prevalence of $Y$ is about 0.5. Here, results will be presented for this particular choice too. However, we shall add that the proportion of configurations for which $\Delta_{AS}$ is negative while the causal measures $\Delta_{CE}$, $\Delta_{CDE}$ and $\Delta_{CDE|A=1, M=1}$ are positive vary with $\nu$ (a configuration standing for a particular choice for the other parameters involved in our generative model,  $\alpha_A, \alpha_U, \alpha_{AU}, \beta_A, \beta_U, \beta_M, \beta_{AM}, \beta_{AU}, \beta_{UM}$ and $\beta_{AUM}$). 

Lastly, to be consistent with Sperrin et al.'s article, we compute measures of association and causal effects on the odds-ratio scale, rather than on the difference scale. We will denote  the corresponding quantities by $OR_{AS}$, $OR_{Sp}$, $OR_{CE}$, $OR_{CDE}$ and $OR_{CDE|A=1,M=1}$. For instance, 
\begin{align*}
OR_{AS} &= \frac{P(Y=1 | A=1, M=1)/P(Y=0| A=1, M=1)}{P(Y=1 | A=0, M=1)/P(Y=0| A=0, M=1)}\\
OR_{CE}&= \frac{P[Y^{A=1}=1|M=1]/P[Y^{A=1}=0|M=1]}{P[Y^{A=0}=1|M=1]/P[Y^{A=0}=0|M=1]}\\
OR_{CDE}&=\frac{P[Y^{1,1}=1]/P[Y^{1,1}=0]}{P[Y^{0,1}=1]/P[Y^{0,1}=0]}\\
OR_{CDE|A=1,M=1}&=\frac{P[Y^{1,1}=1|A=1 ,M=1]/P[Y^{1,1}=0|A=1 ,M=1]}{P[Y^{0,1}=1|A=1 ,M=1]/P[Y^{0,1}=0|A=1 ,M=1]}\cdot
\end{align*}

\subsection{Results}

\begin{figure}\caption{Causal and ``observable''odds-ratios in the case where $\beta_A = \beta_{AM} = \beta_{AU} =  \beta_{UM} = \beta_{AUM} = 0$, with $\beta_M$ set to either 0 (top row) or 1 (bottom row), and for varying values of the other parameters $\alpha_A, \alpha_U, \beta_U$ and $\alpha_{AU}$ . In each panel, along the $x$ axis, one of these parameters is varied from $-3$ to $3$ (left panel: $\alpha_A$, mid-left panel: $\alpha_U$, mid-right panel: $\beta_U$, right panel: $\alpha_{AU}$), and the other parameters are set to a default value ($1$ for $\alpha_A, \alpha_U, \beta_U$ and $0$ for $\alpha_{AU}$).} \label{fig:Figure2Sp}
\includegraphics[scale=0.24]{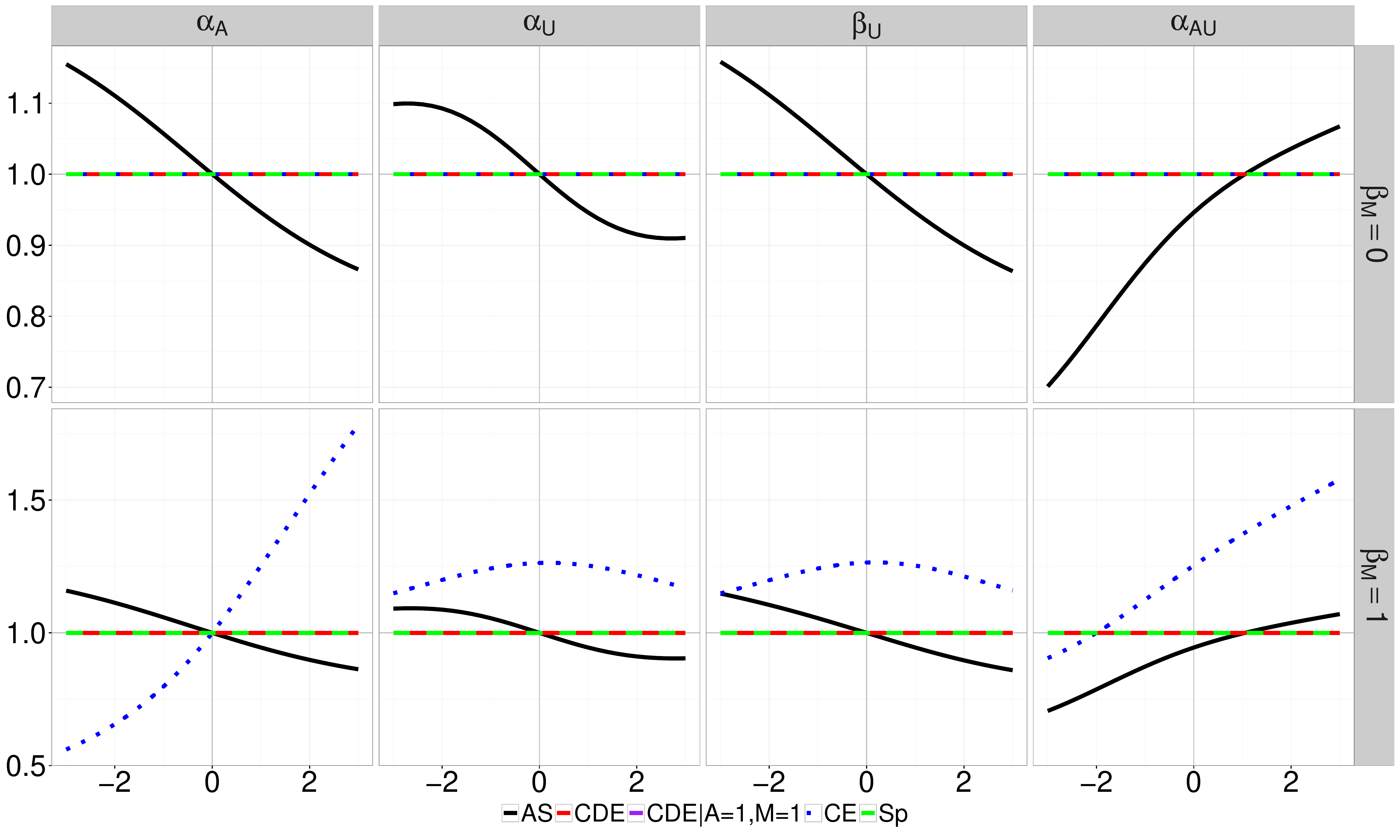}
\end{figure}

We first consider the setting corresponding to Figure 2 of Sperrin et al., where $\beta_A = \beta_{AM} = \beta_{AU} =  \beta_{UM} = \beta_{AUM} = 0$ and $\beta_M=0$ (top row of Figure \ref{fig:Figure2Sp}). We obtain the exact same results as Sperrin et al. Indeed, this setting corresponds to the case where $A$ has neither a direct nor an indirect effet, and causal odds-ratios $OR_{CE}$, $OR_{CDE}$ and $OR_{CDE|A=1,M=1}$ all equal 1. This is one particular situation where $Y^{A=a}\indep A |M$ as mentioned in Section \ref{sec:Error_remarks} above, and therefore $OR_{Sp} = OR_{CE}$ (=1). On the other hand, $OR_{AS}$ is generally not equal to 1, but the difference with the other quantities is typically small.

According to Sperrin al., $\beta_M$ can be set to 0 without loss of generality. The bottom row of Figure \ref{fig:Figure2Sp} shows that it is not the case. Indeed,  when $\beta_M \neq 0$, $OR_{CE}$ is typically different from $1$, as mentioned in Section \ref{sec:Error_remarks}. On the other hand, the quantity $OR_{Sp}$ still equals $1$, and so do $OR_{CDE}$ and  $OR_{CDE|A=1,M=1}$. As for $OR_{AS}$, it behaves as in the case where $\beta_M=0$. This particular case illustrates the discrepancy between the true value of $OR_{CE}$ and the quantity $OR_{Sp}$ studied by Sperrin et al. It further shows that if $\Delta_{CE}$ is the target quantity, then $OR_{AS}$ can be severely biased and, then, that Sperrin et al.'s conclusion is false. In particular when $\alpha_A>2$, $OR_{AS}$ is sensibly lower than $1$ while $OR_{CE}$ is sensibly greater than $1$ (when the other parameters are set to their default values).  

\begin{figure}\caption{Causal and ``observable''odds-ratios in the case where $\alpha_{A} = \alpha_U = 2=\beta_A =\beta_{UM} = 2$, $\beta_U = 3$ and $\beta_{AM} = -2$, for varying values of the other parameters $\alpha_{AU}$, $\beta_{M}$, $\beta_{AU}$ and $\beta_{AUM}$. In each panel, along the $x$ axis, one of these parameters is varied from $-3$ to $3$ (left panel: $\alpha_{AU}$, mid-left  panel: $\beta_M$, mid-right panel: $\beta_{AU}$, right panel: $\beta_{AUM}$) and the other parameters are set to the default value $1$.}\label{fig:Figure2Sp_inter}
\includegraphics[scale=0.24]{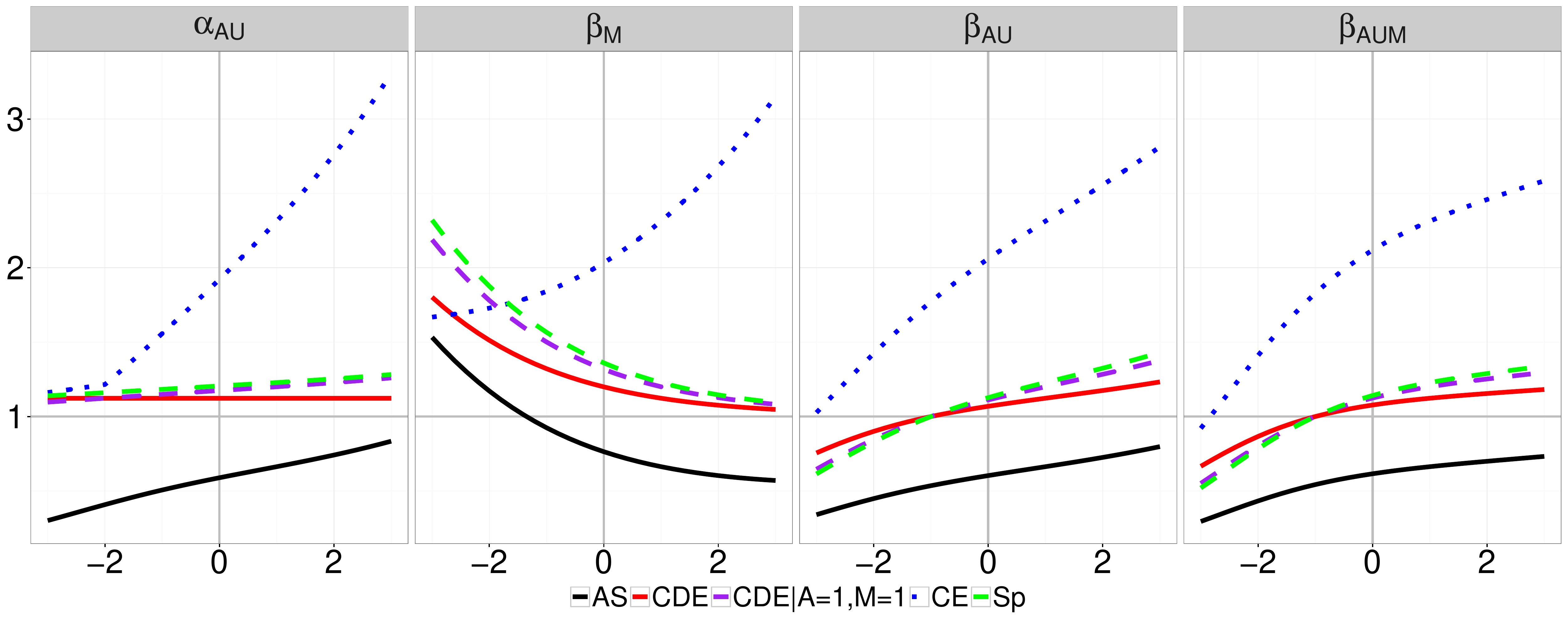}
\end{figure}

However, if the target quantity is $OR_{CDE}$ (or $OR_{CDE|A=1,M=1}$) the bias attached to $OR_{AS}$ is less sensible under the configurations presented on Figure \ref{fig:Figure2Sp}. We present results under configurations where $\alpha_{A} = \alpha_U = \beta_A =\beta_{UM} = 2$, $\beta_U = 3$ and $\beta_{AM} = -2$. In each panel of Figure  \ref{fig:Figure2Sp_inter} one of the four remaining parameters, $\alpha_{AU}$, $\beta_{M}$, $\beta_{AU}$ and $\beta_{AUM}$, varies between $-3$ and $3$, while the other three are fixed at $1$. Overall, under these configurations, $OR_{AS}$ is sensibly inferior to $1$ while  $OR_{CDE}$, $OR_{CDE|A=1,M=1}$ and $OR_{CE}$ are all sensibly superior to $1$. 

To recap, our numerical results establish that a negative association between $A$ and $Y$, when restricting our attention to patients with $M=1$ ($OR_{AS}<1$) does not imply either $OR_{CE}<1$, or $OR_{CDE}<1$ or $OR_{CDE|A=1, M=1}<1$. Therefore, even under the simple generative model considered by Sperrin et al., the ``obesity paradox'' can be artifactual and fully due to collider bias.

\section{Discussion}

Obesity is widely considered as a cause of early death. With the notations used in this article, this means that the causal odds-ratio, $$[P(Y^{A=1}=1)/P(Y^{A=1}=0)] / [P(Y^{A=0}=1)/P(Y^{A=0} = 0)],$$ is superior to $1$. However, several observational studies reported an observed odds-ratio among individuals with diabetes or heart failure  \cite{Anker_2011, Carnethon, Oreopoulos} less than one, suggesting that $OR_{AS}<1$. This observation could indeed be considered as paradoxical if the observed odds-ratio among individuals with chronic disease was a consistent estimate of $OR_{CE}$. However, this is not the case because $M$ is a descendent of $A$, as already suggested in the literature \cite{Lajous_2014}. Contrary to what Sperrin et al. reported \cite{Sperrin}, we show that this difference between $OR_{AS}$ and $OR_{CE}$ can be important even under the simple causal model they considered. In addition, we show that $OR_{AS}$, $OR_{CDE}$ and $OR_{CDE|A=1, M=1}$ share some similarities, but are still different. In particular, by considering additional interaction terms in the generative model proposed by Sperrin et al., we exhibited  configurations where $OR_{AS}<1$ while $OR_{CDE}>1$ and $OR_{CDE|A=1, M=1}>1$ (and $OR_{CE}>1$). Therefore, estimates of $OR_{AS}$ should be regarded with caution since they can not be related to any meaningful causal effects. Furthermore, the confounder $U$ has to be observed in order to estimate the causal quantities $OR_{CDE}$ and $OR_{CDE|A=1, M=1}$. As for $OR_{CE}$, it can not be identified from the distribution of $(A, M, U, Y)$ without further assumptions on the causal model. 

We shall add that even if we could estimate $OR_{CDE}$ and $OR_{CE}$, these quantities might not be appropriate to answer the question of whether weight loss would be beneficial for an obese patient with diabetes or heart failure \cite{Anker_2011}. The risk of early death for such a patient is $P(Y=1 | A=1, M=1) = P(Y^{A=1}=1| A=1, M=1)$, by consistency. But what would be his risk after a weight loss? If it can be assumed that his risk would be the one he would have had in the counterfactual world $\Omega^{A=0}$ (that is, his risk had he never been obese), then it is simply $P(Y^{A=0}| A=1, M=1)$. Because $P(Y^{A=0}=1 | A=1, M=1) = P(^{A=0, M=M^{A=0}}=1 | A=1, M=1)$, it is noteworthy that this assumption implies that this patient might be cured of diabetes after his weight loss.  Under this assumption, the quantity of interest is therefore $P(Y=1 | A=1, M=1) - P(Y^{A=0}=1 | A=1, M=1) = P(Y^{A=1}=1| A=1, M=1) - P(Y^{A=0}=1| A=1, M=1)$. It is related to $\Delta_{CE}$, but also to the excess fraction and, under some assumptions, to the attributable fraction and the probability of disablement \cite{Pearl_2000, Pearl_2009}. 
However, if weight loss is unlikely to cure this patient of diabetes, his risk after a weight loss might rather be  $P[Y^{A=0, M=1} = 1 | A=1, M=1]$. Then, the quantity of interest would be $P(Y=1 | A=1, M=1) - P[Y^{A=0, M=1} = 1 | A=1, M=1] = P(Y^{A=1, M=1}=1| A=1, M=1) - P(Y^{A=0, M=1}=1| A=1, M=1) = \Delta_{CDE|A=1, M=1}$. 

Lastly, because there is no unique and well-defined intervention resulting in weight loss (or weight gain), causal inference on observational data is a particularly complicated task when dealing with obesity  \cite{Hernan_2008}. As a matter of fact, to answer the question whether weight loss would be beneficial for obese patients with diabetes,  a safer roadmap would be first specifying the envisaged intervention(s) which might result in weight loss, and then planning a randomized interventional study.


\section{Technical details}

\subsection{A brief introduction to structural causal models}\label{sec:Details}
\subsubsection{The causal model of Figure \ref{fig:DAG}}
We refer to \cite{Pearl_2000, Hernan_Robins_Book, SWIGs} for a thorough introduction of causal models and counterfactuals. Here, the fundamental concepts are illustrated in the particular case of the model considered in \cite{Sperrin}. Figure \ref{fig:DAG} represents the Directed Acyclic Graph (DAG) attached to this causal model. In this DAG, only the endogenous variables $U, A, M, Y$ are represented, while the corresponding exogenous variables (or disturbances) $\eps_U$, $\eps_A$, $\eps_M$ and $\eps_Y$ are not. The relationships between the endogenous variables can be fully described by the set of structural equations corresponding to this DAG. There is one such equation for each endogenous variable involved in the DAG. It involves a fixed (but unknown) autonomous function, whose inputs are the parents of the variable, along with its associated exogenous variable. In our example, the set of structural equations is  
\begin{align*}
\begin{cases}
A &= f_A(\eps_A)\\
U &= f_U(\eps_U) \\
M &= f_M(A, U, \eps_M)\\
Y & = f_Y(A, M, U, \eps_Y), 
\end{cases}
\end{align*} 
where, $f_A, f_U, f_M$ and $f_Y$ are the unspecified autonomous functions. In this article, we assume that the exogenous variables (or disturbances) are mutually independent. 

The structural equations are further helpful to precisely define the variables we would have been able to observe had we intervened to fix the value(s) of some variable(s). In particular, in the counterfactual world $\Omega^{A=a}$ that would have followed the intervention $do(A=a)$ \cite{Pearl_2000, Pearl_2009, Pearl_2010}, we would have been able to observe the variables 
\begin{align*}
\begin{cases}
U &= f_U(\eps_U) \\
M^{(a)}&=f_M(a, U, \eps_M)\\
Y^{(a)}&= f_Y(a, M^{A=a}, U, \eps_Y).
\end{cases}
\end{align*}
From these equations, it is clear that consistency holds since, if $A=a$, then $M= f_M(a, U, \eps_M) = M^{A=a}$ and $Y = f_Y(a, M, U, \eps_Y) = f_Y(a, M^{A=a}, U, \eps_Y) = Y^{A=a}$.

Variables we would have been able to observe in the counterfactual world $\Omega^{a,m}$ that would have followed the double intervention $do(A=a, M=m)$ can be defined too:
\begin{align*}
\begin{cases}
U &= f_U(\eps_U) \\
Y^{(a,m)}&=f_Y(a, m, U, \eps_Y).
\end{cases}
\end{align*}

From these definitions, we have $Y^{A=a} = Y^{A=a, M^{A=a}}$ as mentioned in Section \ref{sec:Error_remarks}. Here $Y^{A=a, M^{A=a}}$ can be thought of as the outcome we would have been able to observe in the counterfactual world that would have followed the intervention $A=a$ and $M=M^{A=a}$, that is the counterfactual world where $A$ would have been set to $a$ and $M$ would have been set to whatever values it can get under the distribution of $M^{A=a}$; of course this counterfactual world is exactly $\Omega^{A=a}$. 

Moreover, the distribution of the exogenous variables $(\eps_A, \eps_M, \eps_U, \eps_Y)$ induce well-defined probabilities of events such as $\{Y^{A=a}=0\} \cap \{Y=1\}$ or $\{Y^{A=0}=0\} \cap \{M=1\}  \cap \{A=1\}$, etc. See Section 3.4 in \cite{Pearl_2010}. In particular, the quantity $P(Y^{A=a}=1 | M=1)$, which appears in $\Delta_{CE}$, is well-defined and be computed in closed form once we specify the distribution of the disturbances and the form of the structural functions $f_Y$ and $f_M$; see Section \ref{sec:Details_CE} below for an example.

\subsubsection{The causal model of Figure \ref{fig:DAG_simple}}\label{sec:Details_simple}
We conclude this brief introduction by inspecting the simplified causal model of Figure \ref{fig:DAG_simple}. In this case, $A$ is not a parent of $Y$, and there is no confounder. Then, the set of structural equations becomes
\begin{align*}
\begin{cases}
A &= f_A(\eps_A)\\
M &= f_M(A, \eps_M)\\
Y & = f_Y(M, \eps_Y). 
\end{cases}
\end{align*} 
Accordingly, the counterfactual variables $Y^{A=a}, M^{A=a}$ and $Y^{a,m}$ are defined as
\begin{align*}
\begin{cases}
M^{A=a}&=f_M(a,\eps_M)\\
Y^{A=a}&= f_Y(M^{A=a}, \eps_Y)\\
Y^{(a,m)}&=f_Y(m, \eps_Y).
\end{cases}
\end{align*}

\begin{figure}[t]
\caption{The SWIT resulting from the intervention $do(A=a)$ in the simplified causal model of Figure \ref{fig:DAG_simple}.} 
\label{fig:SWIG} 
\vspace{0.3cm}
\begin{center}
\begin{tikzpicture}[scale=1, auto,swap]
\node[varleft] (A)at(0,0){$A$};
\node[varright] (a)at(0.75,0){$a$};
\node[varcounterfactual] (Y)at(5,0){$Y^{A=a}$};
\node[varCond] (M)at(2.5,0){$M^{A=a}$};
\draw[edge] (a)--(M);
\draw[edge] (M)--(Y);
\end{tikzpicture}
\end{center}
\end{figure}
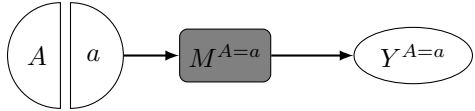

This simplified setting corresponds to that considered in Figure 10 in \cite{SWIGs}. Figure \ref{fig:SWIG} presents the SWIT corresponding to the intervention $do(A=a)$ \cite{SWIGs}. From this representation, it directly follows that $A\indep (Y^{A=a}, M^{A=a})$, and therefore that $Y^{A=a} \indep A| M^{A=a}$; see Section 3.5.3 in \cite{SWIGs}. Then, and as mentioned in Section 3.6.1 in \cite{SWIGs}, the following holds 
$$ \p(Y = 1 | M=1, A=a) = \p(Y^{A=a} | M^{A=a} = 1),$$
which is generally different from $\p(Y^{A=a} | M = 1).$

To better understand why $\p(Y^{A=a} | M^{A=a} = 1) \neq \p(Y^{A=a} | M = 1)$, it is convenient to take a concrete example and compare the (random) sets $\{M^{A=a}=1\}$ and $\{M=1\}$. For instance, consider the generative model described in Section \ref{sec:DataGen}, where we set $\beta_A = \beta_U = \alpha_U =0$ and $\beta_{AM} = \beta_{AU} = \beta_{AUM} = \beta_{UM} = \alpha_{AU} = \alpha_{UM} = 0$ to ensure that there is no arrow pointing from $A$ to $Y$ and no confounder $U$. Then, we have $\{M^{A=0}\} = \{\eps_M \leq {\rm expit}(\alpha_0)\}$, so that $P(M^{A=0}=1) ={\rm expit}(\alpha_0)$, and $\{M^{A=1}\} = \{\eps_M \leq {\rm expit}(\alpha_0 + \alpha_A)\}$, so that $P(M^{A=1}=1) ={\rm expit}(\alpha_0 + \alpha_A)$. As for $\{M=1\}$, we have 
\begin{align*}
\{M=1\} &= \{\eps_M \leq {\rm expit}(\alpha_0 + \alpha_A A)\}\\
&= (\{\eps_M \leq {\rm expit}(\alpha_0 + \alpha_A)\}\cap \{\eps_A \leq p_A\}) \cup (\{\eps_M \leq {\rm expit}(\alpha_0)\}\cap \{\eps_A > p_A\}),
\end{align*}
so that  $P(M=1=1) = p_A {\rm expit}(\alpha_0 + \alpha_A) + (1-p_A){\rm expit}(\alpha_0)$. Therefore,  $\{M=1\} \neq \{M^{A=0}=1\}$, $\{M=1\}\neq \{M^{A=1}=1\}$ and $\{M^{A=0}=1\}\neq \{M^{A=0}=1\}$ as soon as $\alpha_A\neq 0$ in this simplified setting.


\subsection{Derivation of $\Delta_{CE}$ under our generative model}\label{sec:Details_CE}
An analytic formula for $E(Y^{A=a} | M= 1) = P(Y^{A=a} =1 | M=1)$ can be derived from basic probability calculus under the data generation mechanism described in Section \ref{sec:Simul}. For simplicity, we assume that $p_A = p_U = 1/2$. First, 
\begin{align*}
&P(Y^{A=a} =1 | M=1) \\
&= \sum_{\scriptstyle i_A\in\{0,1\} \atop \scriptstyle i_U \in\{0,1\}}  P(Y^{A=a} =1 | M=1, A=i_A, U=i_U) P(A=i_A, U=i_U | M=1 )\\
& = \sum_{\scriptstyle i_A\in\{0,1\} \atop \scriptstyle i_U \in\{0,1\}} P(Y^{A=a} =1 ,  M=1 | A=i_A, U=i_U) \frac{P(A=i_A, U=i_U | M=1 )}{P(M=1 | A=i_A, U=i_U)}\\
& \overset{(*)}{=} \sum_{\scriptstyle i_A\in\{0,1\} \atop \scriptstyle i_U \in\{0,1\}} \int_{0}^{p_M(i_A, i_U)}\frac{P(Y^{A=a} =1  | \eps_M = \eps, A=i_A, U=i_U)}{4P(M=1)} d\eps\\
& = \sum_{\scriptstyle i_A\in\{0,1\} \atop \scriptstyle i_U \in\{0,1\}} \int_{0}^{p_M(i_A, i_U)}\frac{P(\eps_Y \leq p_Y(a, M^{A=a}, U)  | \eps_M = \eps, A=i_A, U=i_U)}{4P(M=1)} d\eps
\end{align*}
where we use the fact that the conditional density of $\eps_M$ given $(A=i_A, U=i_U)$ uniformly equals $1$ over the interval $[0,1]$ to establish equality $(*)$.
Then, successively using the fact that $(i)$  $M^{A=a} = \1[\eps_M \leq p_M(a, U)]$, and $(ii)$ $P(\eps_Y \leq \rho) = \rho$ for any $\rho\in[0,1]$, it follows that 
\begin{align*}
&P(Y^{A=a} =1 | M=1) \\
&= \sum_{\scriptstyle i_A\in\{0,1\} \atop \scriptstyle i_U \in\{0,1\}} \frac{\int_{0}^{[p_M(i_A, i_U)\wedge p_M(a, i_U)]} P(\eps_Y \leq p_Y(a,1,i_U)) d\eps + \int_{[p_M(i_A, i_U)\wedge p_M(a, i_U)]} ^{p_M(i_A, i_U)} p_Y(a,0,i_U) d\eps}{4P(M=1)}  \\
&=\sum_{\scriptstyle i_A\in\{0,1\} \atop \scriptstyle i_U \in\{0,1\}} \frac{p_Y(a,1,i_U)[p_M(i_A, i_U) \wedge p_M(a, i_U)] + p_Y(a,0,i_U) \{p_M(i_A, i_U)- [p_M(i_A, i_U) \wedge p_M(a, i_U)] \}}{4P(M=1)}
\end{align*}

Because $P(M=1) = \sum_{i_A, i_U}p_M(i_A, i_U)/4$, it is straightforward to compute $ P(Y^{A=a} =1 | M=1)$, hence $\Delta_{CE}$, for any combinations of values for the parameters $\alpha_0, \alpha_A, \alpha_U, \alpha_{AU}, \beta_0, \beta_A, \beta_U, \beta_M, \beta_{AM}, \beta_{AU}, \beta_{UM}$ and $\beta_{AUM}$.

\bibliographystyle{ieeetr}

\end{document}